\documentclass[showpacs,twocolumn,pre]{revtex4}
\usepackage{graphicx, amsmath, amsthm, amssymb}
\usepackage{xcolor}

\DeclareUnicodeCharacter{2212}{-}

\begin{document}
  \title{A novel approach to study laser induced void array formation in fused silica }
  
\author{N. Naseri$^1$, M. Crenshaw$^1$, A. Samarbakhsh$^1$, and  L. Ramunno$^2$}
\affiliation{$^1$Departments of Physics and Astronomy, Middle Tennessee State University, Wiser-Patten Science Hall, 422 Old Main Cir, Murfreesboro, TN 37132, U.S.A}
\affiliation{$^2$Department of Physics, University of Ottawa and Nexus for Quantum Technologies Institute, 25 Templeton St., Ottawa, Ontario K1N 6N5, Canada}
\maketitle


\section{introduction}
\begin{figure}[h]
\begin{center}
\includegraphics[width=\linewidth]{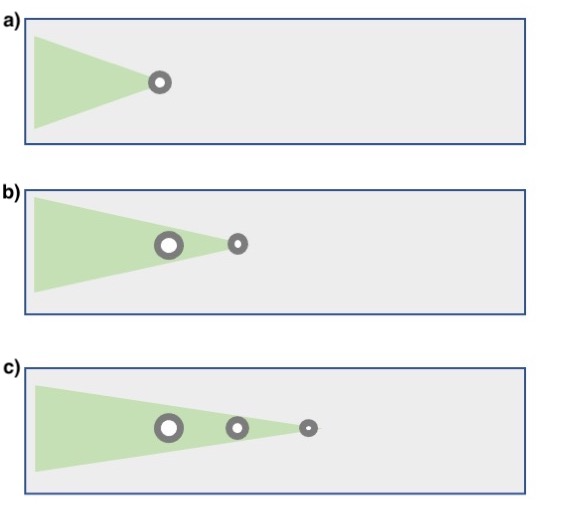}
\end{center}
\caption{Schematic sketch of multi-void formation in silica. a) laser interaction with bulk of fused silica, results in one void structure formation. b) A pre-recorded void is located in the medium, a second void is generated as a result of interaction of the laser pulse with pre-recorded void c)  two pre-recorded voids are located in the medium, a third void is generated as a result of interaction of the laser pulse with pre-recorded voids. This process can continue and eventually multi-voids is generated in fused silica.}\label{fig1p}
\end{figure}

Extensive research spanning several years has been devoted to investigating femtosecond laser-induced modifications and damage in transparent materials. These dedicated efforts have been motivated by the numerous promising applications in the realm of photonics. \cite{Gattas, Zhang2016, Beresna2017, Drev}. These studies explore the complex interactions between femtosecond laser pulses and transparent materials, including polymers, unveiling a diverse array of outcomes influenced by laser intensity and spot size. Of particular interest within this domain is the fascinating phenomenon of laser-induced void array formation. Void array formation in materials has attracted substantial attention due to its potential applications in optical memories, waveguides, gratings, couplers, and the realms of chemical and biological membranes \cite{Graf, Beresna, Bell, Davis, Shimotsuma, Zhang, Zhang2016, Beresna2017, Drev}.
The fabrication of voids within materials such as silica glass \cite{Glezer1996, Glezer1997, Watanabe1999, Watanabe2000, Toratani2005, Dai2016} and polymers \cite{Yamasaki2000, Day2002} has undergone rigorous investigation since the pioneering work of Glezer et al. \cite{Glezer1996}.\\
The morphology of damage structures in transparent materials is determined by the laser focusing conditions. Over the past two decades, extensive research, involving both numerical simulations and experimental investigations of single-pulse filamentation in transparent materials, has elucidated a significant correlation between the interaction of femtosecond lasers with the material and the specific laser focusing parameters. When laser beams are tightly focused, this typically gives rise to the formation of voids \cite{Glezer1997, Schaffer2001}. However, under loose focusing conditions, the predominant outcome often involves the creation of elongated channels characterized by modified refractive indices \cite{Naseri}.
Loose laser focusing is characterized by a laser beam with a spot size that exceeds the laser wavelength by several times. In this scenario, the prominent process at play is the nonlinear Kerr effect, where the interaction between the laser pulse and the material gives rise to the formation of filaments.
Conversely, tight focusing occurs when the laser spot size is either comparable to or smaller than the laser wavelength, making optical focusing the primary mechanism. This tight focusing regime leads to the development of void structures in close proximity to the geometrical focus position \cite{Naseri}.\\
The outcome of the laser-material interaction can also be significantly influenced by the number of laser pulses that interact with the material. Typically, a void structure comprises a central volume of less dense material encircled by a region of higher density \cite{Glezer1996, Glezer1997}. Despite extensive research dedicated to understanding the mechanisms behind multi-void formation in materials, the fundamental aspects and mechanisms governing self-void array formation in dielectrics and polymers remain poorly understood.\\
Previous studies have proposed that self-focusing \cite{Kanehira} and spherical aberration \cite{Song, Wang} play a role in the creation of laser-induced multi-voids. However, it has been observed that the laser re-focusing period \cite{Wu} significantly exceeds the timescale of void array formation. Moreover, self-focusing predominantly occurs under loose laser focusing conditions, which is inconsistent with the formation of multi-voids. Consequently, nonlinear self-focusing is unlikely to be the driving mechanism behind void array formation.
The comprehensive understanding of the mechanisms governing void array formation in materials holds the potential for significant advancements in precision micromachining and improved control over induced material changes in optical engineering. Recently, we showed that the process governing the void array formation in PDMS is linear geometrical focusing.\cite{Naseri2023}. We will follow similar approach to study the mechanism of void array formation in fused silica.\\
In this paper, we present the findings of our three-dimensional, high-resolution finite-difference-time-domain (FDTD) simulations, providing insights into the mechanisms underlying the formation of multi-voids in a dielectric medium when exposed to multiple laser pulses. Our simulations reveal that the initial laser pulse induces a localized modification of the refractive index within the medium  where the first void forms (fig. \ref{fig1p}-a). This modification acts as a focusing microlens, concentrating subsequent laser pulses. As these successive pulses are focused by the pre-recorded void, a second refractive index modification occurs in front of the pre-recorded void(fig. \ref{fig1p}-b). The size of the initial void influences the size of the second void and the spacing between them. This iterative process is reinforced by multiple laser pulses, eventually leading to the formation of an entire array of voids (fig. \ref{fig1p}-c).

Schematic figure \ref{fig1p} (a-c) illustrates the interaction of the laser pulse with the bulk of fused silica (fig. \ref{fig1p}-a), demonstrating the generation of the first void. Figure \ref{fig1p}-b portrays the interaction of the laser with a pre-existing void, resulting in laser light being focused in front of the void and leading to the creation of a second void. Figure \ref{fig1p}-c illustrates the interaction of the laser with two pre-existing voids, resulting in the generation of a third void due to laser focusing in front of the second void. Our model effectively mitigates the impact of spherical aberrations and confirms that the formation of multi-voids is a cumulative effect of multiple laser pulses, and is primarily operating a linear mechanism. 

\section{Numerical method}
To understand the mechanism of multi-void formation, we performed three-dimensional high resolution finite-difference-time-domain (FDTD) simulations. To date, such simulations relied on nonlinear pulse propagation with plasma generation and dynamics, using approximations such as slowly varying envelopes or evolving waves, to obtain an evolution equation for the pulse \cite{Cou}. 
While these approaches have proven effective in certain cases when the right approximations are made for the specific problem, such as in cases where the focusing is not too tight, the medium is relatively uniform, and the plasma created is not too dense \cite{Dub}. However, since the spot size of the laser and the void size in materials are usually a fraction of the laser wavelength, a more rigorous computational approach that does not make any assumptions about light propagation is required \cite{Popov2011, Bulga, Rudenko}.\\
Initially, we examine the interaction of a single laser pulse with fused silica. Our numerical modeling shows that increasing the laser energy leads to the formation of elongated single voids rather than a void array. Building upon this, we then modeled the first damage structure (void) as pre-existing in the medium and investigate the interaction of subsequent laser pulses with this pre-recorded void. We observe the generation of a second void as a consequence of field enhancement in front of the first void. Subsequently, we consider the scenario where the two generated voids are pre-recorded in the medium, leading to the generation of a third void. Our method reveals that this process can continue, enabling the generation of successive voids through multi-pulse laser interaction in fused silica. In our 3D numerical simulation model, Maxwell's equations are solved using the FDTD method \cite{Taflove2005} via the Yee algorithm \cite{Yee,Popov} with constitutive  relations (cgs units), $
  \textbf{H}=\textbf{B},
~
  \textbf{D}=(1+4\pi(\xi_l+\xi_kE^2)\textbf{E}$
and  current density
  $\textbf{J}=\textbf{J}_p+\textbf{J}_{PA}$. $\textbf{E}$ and $\textbf{B}$ are the electromagnetic fields, $\textbf{D}$ the displacement vector, $\textbf{H}$, the magnetic field auxiliary vector, $\xi_l$ is the linear susceptibility of the material, and $\xi_k$ is the Kerr susceptibility, which we take to be constant. The electromagnetic response of the generated plasma is represented by $\textbf{J}_p$ and  laser depletion due to photo ionization (PI) by  $\textbf{J}_{PA}$. 
  \begin{figure}[h]
\begin{center}
\includegraphics[width=\linewidth]{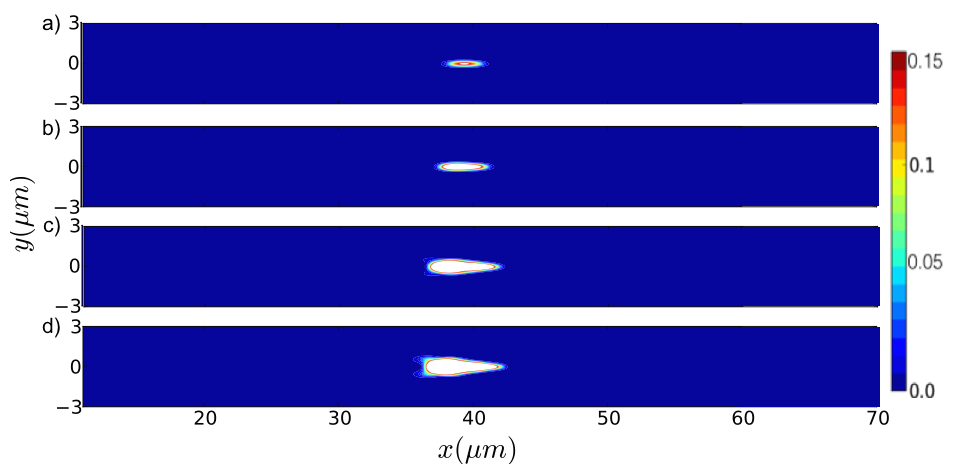}
\end{center}
\caption{Contour plots of electron density
 for intensities of $I=5\times10^{13}$
 (a), $1.1\times 10^{14}$
 (b), $3.3\times10^{14}$
 (c), and $4.9\times10^{14}~ W/cm^2$
 (d).  }\label{fig2}
\end{figure}\\
 The evolution of the free electron density, $n$,  is described by: $ \frac{d n}{d t}=W_{PI}(|\varepsilon|)$, where $W_{PI}$ is the PI rate. Following Keldysh's formulation \cite{Keldysh} for the PI rate $W_{PI}$, the adiabaticity parameter for solids is $\gamma=\omega_0\sqrt{m_{eff}U_i}/eE$, where $m_{eff}=0.635m_e$ denotes the reduced mass of the electron and the hole, $U_i$ is band gap energy, $E$ is the laser electric field, and $\omega_0$ is the laser frequency.
 While we have implemented a model of avalanche ionization that follows Ref. \cite{Rethfeld},
we find that our simulation results for $50~ fs$ pulses including avalanche ionization were identical
to equivalent simulations that did not include avalanche ionization. Thus, to save computational
resources, we did not enable avalanche ionization in our code for the simulations presented here.
We assume a laser beam focused by a perfectly reflecting parabolic mirror characterized by
a given $f \#$, corresponding to laser beam waist size of $w_0 = 0.69~ \mu m$ at the focus in free space. Our model thereby eliminates the effects of spherical aberrations. The laser beam
incident onto the mirror is a Gaussian beam whose waist is half the size of the mirror.
To describe the fields focused by the parabolic mirror, the Stratton-Chu integrals \cite{Popov2009,Stratun} are
used, which specify the exact electromagnetic field emitted by the given parabolic surface. This
field is calculated on five boundaries of the 3D FDTD simulation in a total field/scattered field
framework. The laser pulses are Gaussian in
time with a pulse duration of $50 ~ fs$ and a wavelength of $\lambda = 800~ nm $ and they are linearly
polarized along the y direction and propagating along the x direction. The geometrical laser
focus is located at $x = 40 \mu m$, and the simulation domain is $60~\mu m \times 16~ \mu m \times 16~ \mu m$, with
grid size $\Delta x = 0.02$ and $\Delta y = \Delta z = 0.016 ~\mu m$. We chose this domain size to insure that, in all simulations, the laser pulse had ceased creating plasma well before it exited
the domain. 
\begin{figure}[h]
\begin{center}
\includegraphics[width=\linewidth]{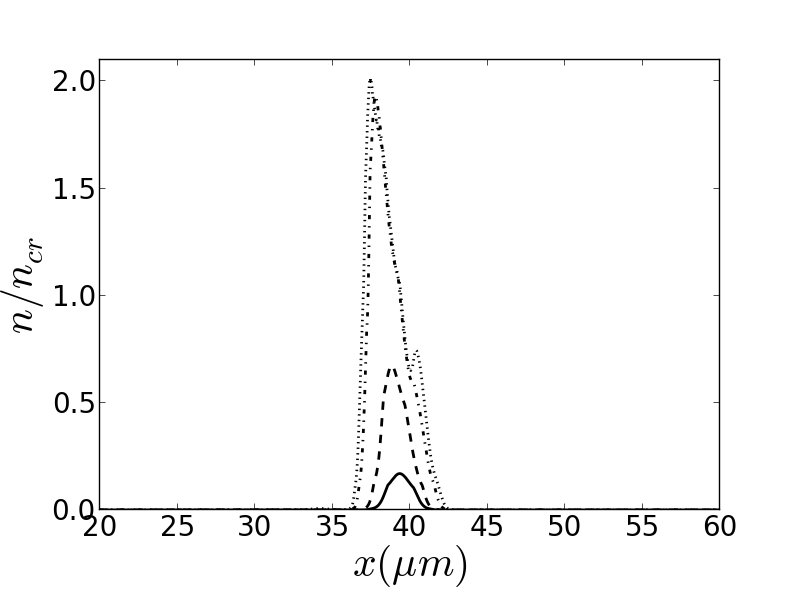}
\end{center}
\caption{On-axis electron density along the laser propagation direction $x(\mu m)$ from simulations for $w_0 = 0.69~ \mu m$ for $I = 5\times 10^{13}$ (solid line), $1.1 \times 10^{14}$  (dashed line), $3.3 \times 10^{14}$ (dashed-dotted line), and $4.9\times 10^{14}$ (dotted line) $W/cm^2$. }\label{fig3}
\end{figure}
Further, we ran simulations where we placed the geometric focus much deeper
within the simulation domain and found no difference in our results. 
The linear refractive index of fused silica is $1.45$. the band gap energy is $9~ eV$, the third order nonlinear susceptibility is $\chi^3 = 1.9 × 10^{-4}~ esu$, and the saturation density is $10~n_{cr}$, where $n_{cr} =\frac{m_e\omega_0^2}{4\pi^2}$ is the critical plasma density; and $\omega_0$ refers to laser frequency.
The saturation density is estimated at $n_s \approx 10 n_{cr}$, where $n_{cr} \approx 1.75 \times 10^{21}~ cm^{−3}$ is the critical electron density for the free-space laser wavelength  $0.8 ~ \mu m$. The saturation density $n_s$ is the maximum density that can be reached when every molecule is ionized. 
The FDTD  simulates linear and nonlinear laser propagation in the medium, plasma generation, ionizational and collisional energy losses as well as laser interaction with the created plasma.
\begin{figure}[h]
\begin{center}
\includegraphics[width=\linewidth]{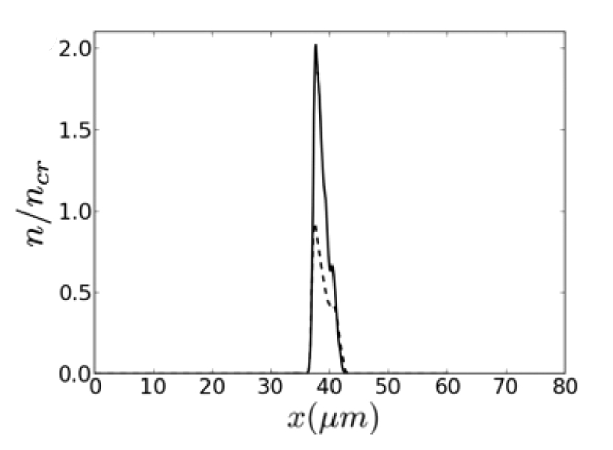}
\end{center}
\caption{On-axis final electron density for 
 $w_0=0.69~\mu m$.  The solid line correspond to the simulations of Fig. 2-c, whereas the dashed lines correspond to equivalent simulations without the Kerr effect. }\label{fig4}
\end{figure}
\section{Simulation results and discussion}
\subsection{Laser interaction with bulk of Fused silica}
The mechanism of void array formation, as observed in the experiments is a multi-laser pulse effect. In order to understand what precisely drives multi-void formation, we first performed simulations of laser interaction with bulk of fused silica. The first laser pulse is expected to generate the first void structure in the bulk of silica. 
In what follows we considered how the input laser energy for fixed laser spot size (and fixed laser pulse duration) affects the interaction with bulk fused silica. Figure \ref{fig2} shows electron density contour plots after the laser pulse left the medium, for peak incident laser intensities of: $5 \times 10^{13}, ~1.1 \times 10^{14},~ 3.3 \times 10^{14}, ~4.9 \times 10^{14} ~W/cm^2$ (Fig. \ref{fig2}(a)-d, respectively). We observe that when the laser peak intensity is varied, the focus position does not change considerably. We found that the threshold intensity for permanent damage happens for an incident laser peak intensity of $5 \times 10^{13}~ W/cm^2$ which leads to a plasma size of $\approx 1~ \mu m$, with peak plasma density $0.16n_{cr}$. Since here geometrical focusing is very strong, there is not as much time for Kerr self-focusing to build. Thus, a higher incident intensity is required to reach the damage threshold in the laser interaction region for very tight focusing than would be required for looser focusing. Increasing the laser intensity to $1.1 \times 10^{14}~ W/cm^2$ leads to a longer ($\approx 2.9~ \mu m$) oval shape structure with maximum electron density $0.65n_{cr}$. The damage area for laser peak intensity of $3.3 \times 10^{14}~ W/cm^2$ is further elongated ($\approx 4.6 ~\mu m$) and has a pear shape structure the same as increasing the peak laser intensity to $4.9 \times 10^{14}~ W/cm^2$ leads to very similar structure as $3.3 \times 10^{14}~ W/cm^2$. Figure 3 shows the on-axis values of the electron densities corresponding to the simulations of Figure 2. While the plasma shape elongates as the intensity is increased, and the plasma density increases, we see a saturation in the electron density for $3.3 \times 10^{14}~ W/cm^2$ and above.
\begin{figure}[h]
\begin{center}
\includegraphics[width=\linewidth]{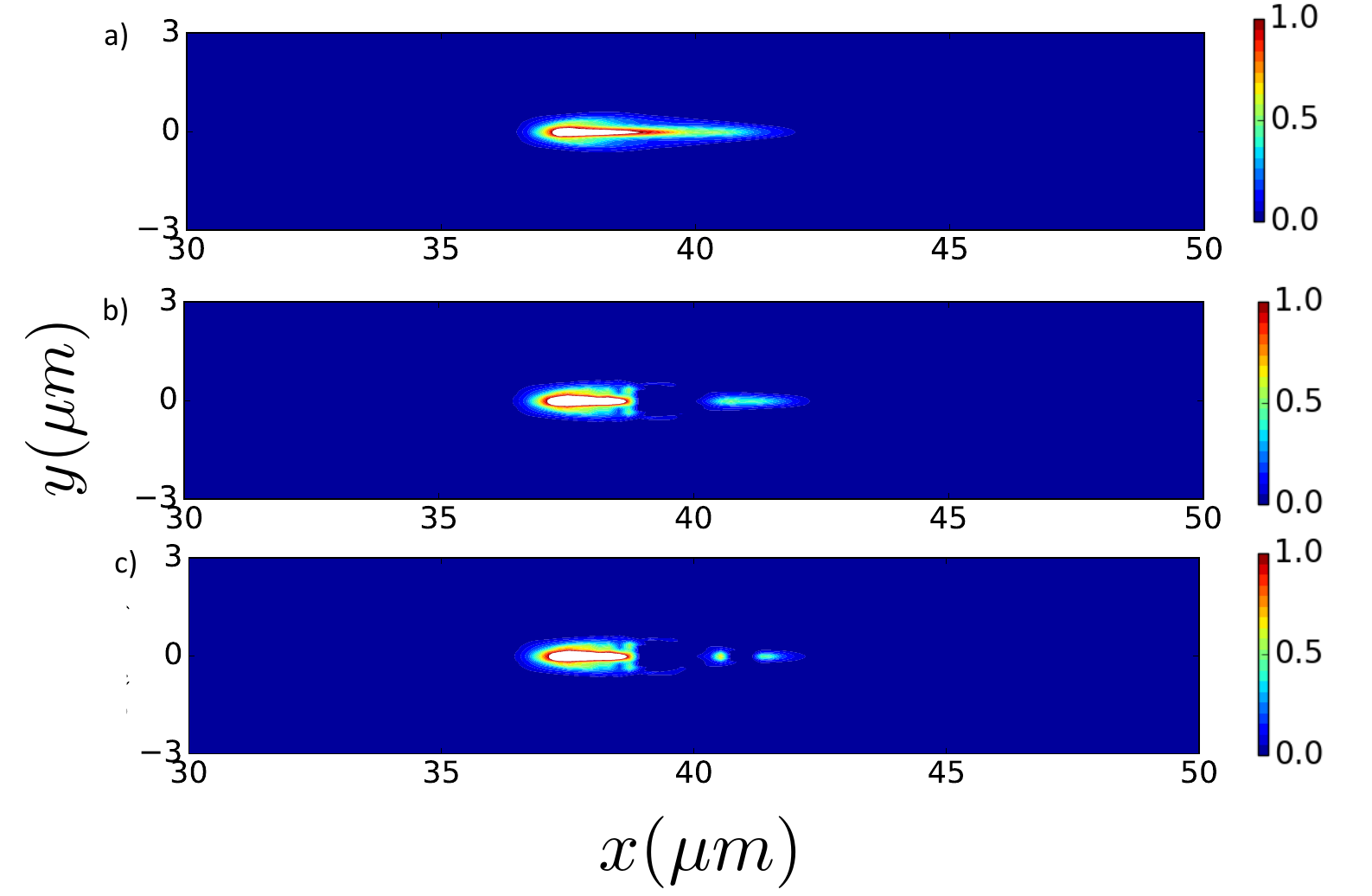}
\end{center}
\caption{Contour plot of final electron density normalized to $n_{cr}$, (a) laser interaction with bulk of silica, (b) with one pre-recorded void located at $x=37.3~\mu m$, and (c) with two pre-recorded voids located at $x=37.3, 41.0~\mu m$. }\label{fig5}
\end{figure}
The magnitude of the change of refractive index $\Delta n$
 corresponding to permanent damage in fused silica has been measured in the experiments in Ref. \cite{Couairon2005}. They found that permanent damage happens in fused silica when $n>0.15n_{cr}$. Thus, as a rough indication of the permanent damage zones predicted by our simulations, we indicate in white in Fig. \ref{fig1} (and subsequent figures) the regions in the electron density contour plots corresponding to $n>>0.15n_{cr}$. 
The resulting electron density structure, shown in Figure \ref{fig2}, closely resembles the experimental findings, featuring an oval-shaped void. The simulation also indicates that the laser focus position is near the geometrical focus, suggesting that Kerr nonlinearity is not the primary interaction mechanism. The plasma is confined to near the geometric focus, and is rapidly formed. Plasma defocusing is also seen, but since geometric defocusing is so strong (after the geometrical focus), it dominates over Kerr self-focusing. \\
\begin{figure}[h]
\begin{center}
\includegraphics[width=\linewidth]{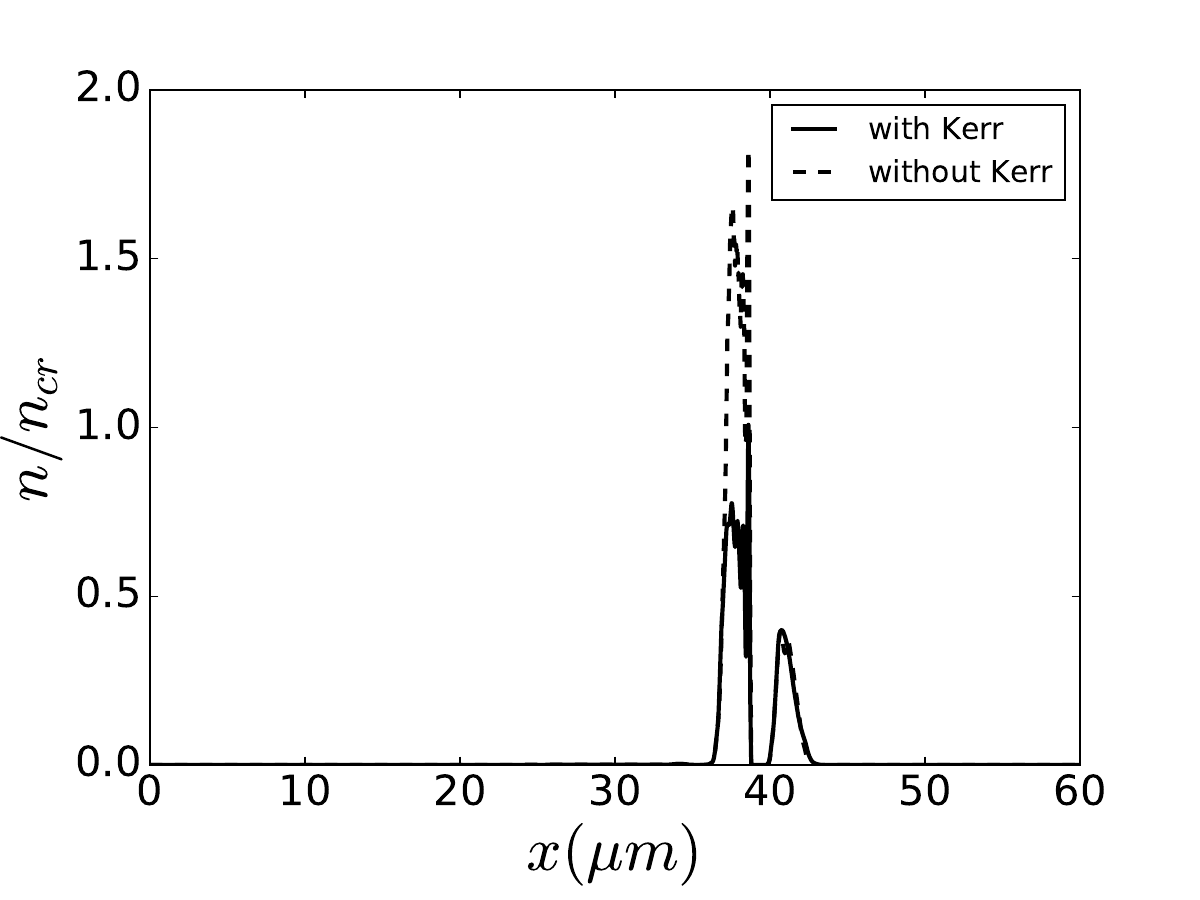}
\end{center}
\caption{On axis electron density normalized to $n_{cr}$ with and without  Kerr effect  with one pre-recorded void located at $x=39.3~\mu m$. Note that even without Kerr effect refocusing, electron density still exceeds $0.15n_{cr}$ }\label{fig6}
\end{figure}
To investigate the role of Kerr self-focusing, we conducted simulations with the nonlinear Kerr effect disabled by setting the Kerr susceptibility to zero, while all other parameters stayed the same. The results, shown in fig. \ref{fig4}, confirm that Kerr nonlinearity is not the dominant mechanism, and that geometric focusing plays a more significant role. Figure 4 displays the plasma density along the laser axis after the laser pulse exits the medium, with the solid and dash-dotted curves representing the results obtained with and without the Kerr effect, respectively. As can be seen in the figure, the length of the void structure consistent. Plasma density is higher when Kerr nonlinearity is included; however, it is important to note that the plasma density exceeds the threshold density in both simulations. The simulation confirms that the mechanism of single void formation in fused silica is a linear process. In next section, we will use the results of this section for laser peak intensity of $1.1 \times 10^{14}~ W/cm^2$, to study multi-void formation in fused silica.
\section{Laser pulse interaction with one and two pre-recorded voids in fused silica }
We have developed a novel model \cite{Naseri2023} to better understand the mechanism behind the formation of multiple voids observed in the experiments \cite{Toratani2005}. For a subsequent simulation representing a second laser shot, we embed the first void  produced by the first pulse in the bulk of fused silica. This consists of a concentric sphere with an inner radius of $0.4~\mu$m and an outer radius of $0.5~\mu$m. The center of the sphere was filled with air to simulate the refractive index of the void, while the shell had a slightly higher linear refractive index than fused silica to model a shell of increased density ($\Delta n=0.1$). The location of the modeled void was determined based on the location where the maximum plasma density was observed in the previous simulation with the Kerr effect included $(x=37.3~\mu$m) (fig. \ref{fig5}-a). All simulations with voids are performed with laser peak intensity of $1.1 \times 10^{14}~ W/cm^2$, while other parameters were kept the same.
\begin{figure}
\begin{center}
\includegraphics[ width=0.5\textwidth]{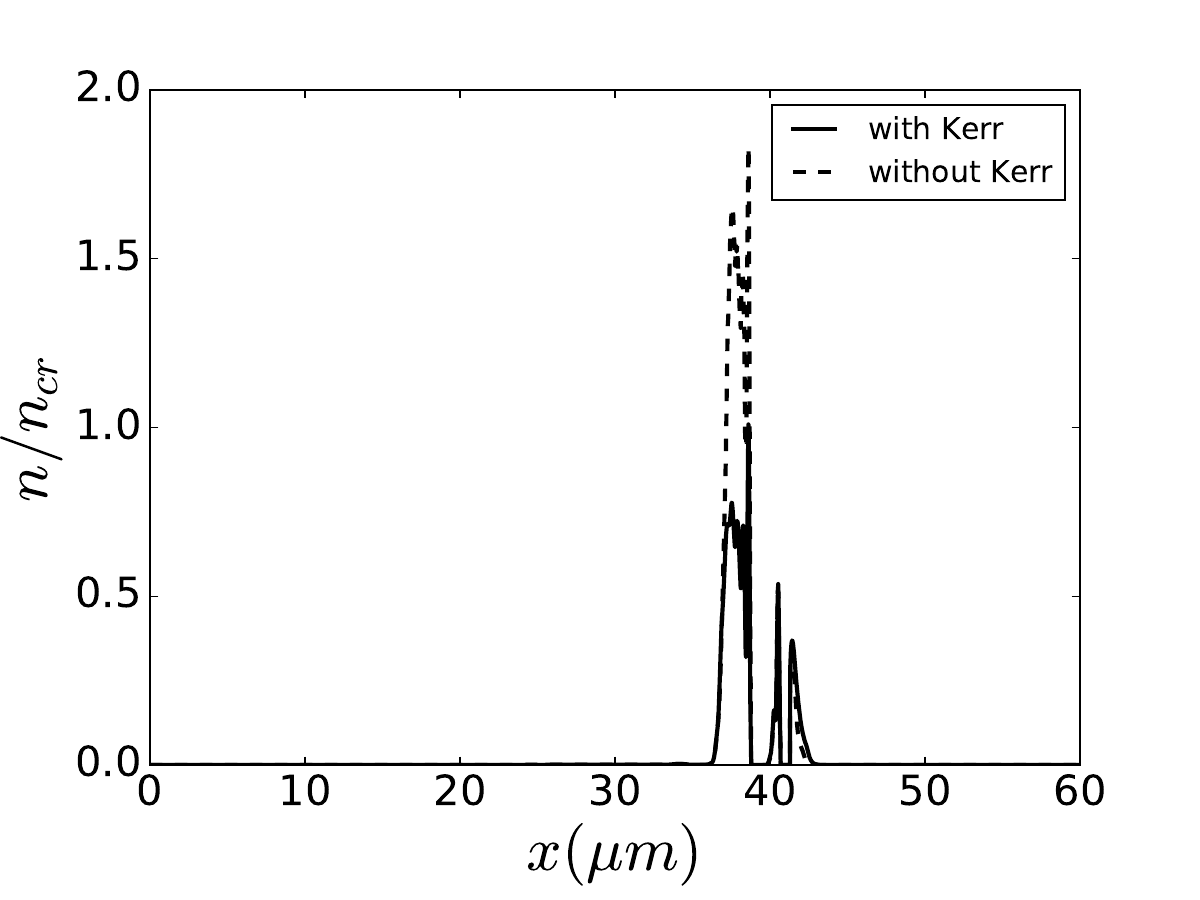}
\end{center}
\caption{ Lineouts of on axis plasma density normalized to $n_{cr}$ in the middle $x-y$ plane. Solid and dashed curves corresponds to simulations without Kerr nonlinear  and with nonlinear Kerr effect, respectively. Two spherical voids are located at $x=37.3, 41.0 ~\mu m$.}\label{fig7}
\end{figure}
The subsequent laser pulse, which could be the second or third pulse in the experiment, enters the medium from the left boundary and interacts with the medium containing the pre-recorded void. We see that electric field enhancement occurs to the right of the pre-recorded void, leading to the formation of a second void. The plasma density contour plot after the laser pulse has left the medium is shown in fig. \ref{fig5}-b. A portion of the laser pulse scatters from the void and focuses in front of it, at a distance of $3.3~\mu m$ from the center of the void, where the laser intensity is sufficient to ionize the fused silica medium, resulting in the formation of a second void at a distance of $40.6~\mu$m with a radius of $\approx 0.3~\mu m$. Furthermore, we find plasma density above threshold at the perimeter of the pre-existing void, suggesting that void sizes grow with increasing number of shots, as  observed experimentally  (see figs. \ref{fig2},\ref{fig3}).\\
To verify that the void generation continued with subsequent laser pulses, we implemented pre-recorded voids, both the first and the second voids, as concentric spheres with refractive index of air in the centers, and densified shells with slightly higher linear refractive index than the medium ($\Delta n=0.1$). The pre-recorded voids were placed at the locations where the first $(x=37.3~\mu m)$ and the second void $(x=40.6~\mu m)$ were observed (fig.\ref{fig5}). Figure \ref{fig5}-c illustrates the interaction of the laser with two voids previously recorded.  The contour plot of the electron density shows the generation of the third void in front of the second void, where field enhancement occurred in the right side of the second void. The radius of the third void was $\approx 0.25~\mu m$, and the electron density was above the damage threshold. The process of void array generation in Fused silica can continue by adding the voids one after another using the method presented here.\\
To understand the mechanism of void formation, we repeated the simulations with one pre-recorded void but this time we turned off the Kerr nonlinearity. Figure \ref{fig6} shows the on axis electron density comparing the two simulations. We can see that the second void is generated in both simulations at the same location and has the same maximum electron density. The electron density on the sides of the pre-recorded void is higher in simulation without including Kerr effect. We conclude that Kerr nonlinearity is not the leading mechanism in second void formation. Therefore the second void formation is not a nonlinear mechanism.
The mechanism presented in this study clearly illustrates the successive generation of voids in fused silica as a result of multi-laser pulse interaction. To confirm that nonlinearity is not the leading process in void generation, we conducted simulations with one and two voids while turning off the Kerr nonlinearity and keeping other parameters the same as in Fig. \ref{fig5}-c.  The on-axis lineouts of electron densities in fig. \ref{fig7} (solid curves) indicate that the second and third voids are located very close to the simulation result including Kerr effect (dashed curves). Additionally, the maximum electron densities at the position of the second and third voids are still above the damage threshold. Therefore, the mechanism is primarily linear. These results ultimately compared well to structures recorded by multi-laser interaction with fused silica in  the literature\cite{}.
\section{Conclusion}
This study investigated multi-void formation in fused silica using high resolution Finite-Difference-Time-Domain (FDTD) simulations. 
First we studied laser interaction with bulk of fused silica. The threshold laser intensity for single void structure formation in fused silica with laser pulse duration of $50~fs$, wavelength of $800~nm$, and laser spot size of $0.69~\mu m$ is $5\times 10^{13}~W/cm^2$, which corresponds to laser energy of $1.53~\mu J$
To understand the mechanism of multi-void formation, we 
modeled a pre-recorded void in the material as concentric sphere with densified shells and simulated the laser's interaction with the void, this demonstrated that a second void is generated as a result of laser pulse interaction with pre-recorded void. To confirm the mechanism, we carried out another simulation with two pre-recorded voids as concentric spheres with densified shells and simulated this further interaction with the voids resulted in the third void formation behind the second pre-recorded void. To understand the mechanism, we repeated the simulations with Kerr nonlinearity terms turned off in the simulations. Comparing the results of simulations with and without Kerr effect revealed that the mechanism of multi-void formation  in fused silica is a linear mechanism. 
This study provides valuable insight into the mechanism behind the formation of void arrays in fused silica. The simulation results agree  with the experimental results to further validate the model and gain a better understanding of the physical processes involved in the generation of void arrays in fused silica.



\begin{thebibliography}{27}
\bibitem{Gattas}
R. Gattas, E. Mazur, Nat. Photonics 2 (4) 219-225 (2008).
\bibitem{Zhang2016}
J. Zhang, A. Čerkauskaitė, R. Drevinskas, A. Patel, M. Beresna, and P. G. Kazansky, Proc. of. SPIE, 9736, 9736U (2016).

\bibitem{Beresna2017}
 M. Beresna, M. Gecevičius, and P. G. Kazansky,  Adv. Opt. Photonics 6(3), 293 (2014).
\bibitem{Drev}
 R. Drevinskas, M. Beresna, J. Zhang, A. G. Kazanskii, and P. G. Kazansky, Adv. Opt. Mater. 5(1), 1600575 (2017). 
\bibitem{Graf}
R. Graf, A. Fernandez, M. Dubov, H. J. Brueckner, B. N. Chichkov, and A. Apolonski,  Appl. Phys. B: Lasers Opt. 87(1), 21–27 (2007).
\bibitem{Beresna}
M. Beresna, M. Gecevičius, and P. G. Kazansky,  Opt. Mater. Express 1(4), 783–795 (2011). 
\bibitem{Bell}
 Y. Bellouard and M.-O. Hongler, Opt. Express 19(7), 6807–6821 (2011).
\bibitem{Davis}
K. M. Davis, K. Miura, N. Sugimoto, and K. Hirao,
Opt. Lett. 21(21), 1729–1731 (1996)
\bibitem{Shimotsuma}
Y. Shimotsuma, M. Sakakura, P. G. Kazansky, M. Beresna, J. Qiu, K. Miura, and K. Hirao, Adv. Mater. 22(36), 4039–4043 (2010).
\bibitem{Zhang}
 J. Zhang, M. Gecevičius, M. Beresna, and P. G. Kazansky,  Phys. Rev. Lett. 112(3), 033901 (2014).
\bibitem{Glezer1996}
  E. N. Glezer, M. Milosavljevic, L. Huang, R. J. Finlay, T.-H. Her, J. P.
Callan, and E. Mazur, Opt. Lett. 21, 2023 (1996).
\bibitem{Glezer1997}
 E. N. Glezer and E. Mazur, Appl. Phys. Lett. 71, 882 (1997).
\bibitem{Watanabe1999}
M. Watanabe, S. Juodkazis, H. Sun, S. Matsuo, H. Misawa, M. Miwa, and
R. Kaneko, Appl. Phys. Lett. 74, 3957 (1999).
\bibitem{Watanabe2000}
M. Watanabe, S. Juodkazis, H. Sun, S. Matsuo, and H. Misawa, Appl.
Phys. Lett. 77, 13 (2000).
\bibitem{Toratani2005}
E. Toratani, M. Kamata, and M. Obara, App. Phys. Lett. 87, 171103 (2005).
\bibitem{Dai2016}
Y. Dai, P. J. Song, M. Beresna, and P. G. Kazansly, Opt. Exp. 24, 19344-19353 (2016).
\bibitem{Yamasaki2000}
K. Yamasaki, S. Joudkazis, M. Watanabe, H.-B. Sun, S. Matsuo, and H. Misawa, Appl. Phys. Lett. 76, 1000 (2000).
\bibitem{Day2002}
D. day, M. Gu, Applied Phys. Lett. 80, 2404 (2002).
\bibitem{Schaffer2001}
38. C. B. Schaffer, A. Brodeur, J. F. Garcia, and E. Mazur,  Opt. Lett. 26(2), 93–95 (2001).
\bibitem{Naseri}
N. Naseri, G. Dupras, L. Ramunno, Opt. Exp. 28,26977 (2020).
\bibitem{Kanehira}
S. Kanehira, J. Si, J. Qiu, K. Fujita, and K. Hirao, Nano Lett. 5(8), 1591–1595 (2005).
\bibitem{Song}
J. Song, X. Wang, X. Hu, Y. Dai, J. Qiu, Y. Cheng, and Z. Xu,  Appl. Phys. Lett. 92(9), 092904 (2008).
\bibitem{Wang}
X. Wang, F. Chen, Q. Yang, H. Liu, H. Bian, J. Si, and X. Hou,  Appl. Phys. A 102(1), 39–44 (2011).
\bibitem{Wu}
Z. Wu, H. Jiang, L. Luo, H. Guo, H. Yang, Q. Gong,  Opt. Lett. 27, 448 (2002).
\bibitem{Cou}
A. Couairon, E. Brambilla, T. Corti, and T. et al.,  Eur. Phys. J.: Spec. Top. 199(1), 5–76 (2011).
\bibitem{Dub}
A. Dubietis and A. Couairon,  SpringerBriefs
in Physics. (Springer, Cham), (2019).
\bibitem{Popov2011}
  Popov K I, McElcheran C, Briggs K, Mack S and Ramunno L  Opt. Express, 19, 271 (2011).
  
\bibitem{Bulga}  
N. M. Bulgakova, V. P. Zhukov, Y. P. Meshcheryakov, L. Gemini, J. Brajer, D. Rostohar, and T. J. Mocek,  J. Opt.
Soc. Am. B 31(11), C8–C14 (2014).
\bibitem{Rudenko}
A. Rudenko, J.-P. Colombier, and T. E. Itina,  Phys. Rev. B 93(7), 075427 (2016).
\bibitem{Taflove2005}
  A. Taflove, S. C. Hagness, \textit{Computational Electrodynamics, 3rd. ed. 2005)}
  \bibitem{Yee}
  K. S. Yee, IEEE Trans. Antennas Propag. \textbf{AP 14}, 302 (1966).
  \bibitem{Popov}
  K. I. Popov, C. McElcheran, K. Briggs, S. Mack, and L. Ramunno, Opt. Express, 19, 271 (2010).
\bibitem{Keldysh}
L. V. Keldysh, Sov. Phys. JETP \textbf{2}, 1307 (1965).
\bibitem{Rethfeld}
B. Rethfeld,  Phys. Rev. Lett. 92(18),
187401 (2004).
\bibitem{Popov2009}
K. I. Popov, V. Yu. Bychenkov, W. Rozmus, R. D. Sydora, and S. S. Bulanov,  Phys. Plasmas 16(5), 053106 (2009).

 \bibitem{Stratun}
 J. A. Stratton and L. J. Chu, Phys. Rev. \textbf{56}, 99 (1939).
 \bibitem{Couairon2005}
 A. Couairon, L. Sudie, M. Franco, B. Prade, and A. Mysyrowicz, “Filamentation and damage in fused silica induced by tightly focused femtosecond laser pulses,” Phys. Rev. B 71(12), 125435 (2005).
  
  \bibitem{Damore2004}
  F. D’Amore , M. Lanata, S. M.  Pietralunga, M. C. Gallazzi, and G. Zerbi,   Opt. Mater. 24 661–5 (2004).
  

 \bibitem{Naseri2023}
  N. Naseri , A. Alshehri, L. Ramunno , and R. Bhardwaj, arXiv:2305.02976, (2023).

\bibitem{Bloem1997}
N. J. Bloembergen,  Nonlin. Opt. Phys. Mater. 6 377 (1997).







\end{thebibliography}
\end{document}